\documentclass{article}
\usepackage[utf8]{inputenc}
\usepackage{hyperref}
\usepackage{amsmath}
\usepackage{amsfonts}
\usepackage{amssymb}
\usepackage{graphicx}
\usepackage{float}
\usepackage{enumitem}
\usepackage{multicol}
\usepackage{natbib}
\usepackage[font=small]{caption}
\usepackage{subfigure}

\usepackage[a4paper,left=1.50cm, right=1.50cm, top=1.50cm, bottom=1.50cm]{geometry}

\author{}

\title{SKETCHING 1-D STABLE MANIFOLDS OF 2-D MAPS WITHOUT THE INVERSE}

\begin{document}
	
	\begin{center}
		{\Large \textbf{Sketching 1-D Stable Manifolds of 2-D Maps without the inverse}}\\
		\vspace{1.5em}
		{\large Vaibhav Ganatra}\\
		\textit{Department of Computer Science and Information Systems\\
Birla Institute of Technology and Science Pilani, K.K. Birla Goa Campus, India\\
f20190010@goa.bits-pilani.ac.in}\\
        \vspace{1em}
		{\large Soumitro Banerjee}\\
		\textit{Department of Physical Sciences\\
Indian Institute of Science Education and Research, Kolkata, India\\
soumitro@iiserkol.ac.in}
	\end{center}
	
	\begin{abstract}
	\noindent Saddle fixed points are the centerpieces of complicated dynamics in a system. The one-dimensional stable and unstable manifolds of these saddle-points are crucial to understanding the dynamics of such systems. While the problem of sketching the unstable manifold is simple, plotting the stable manifold is not as easy. Several algorithms exist to compute the stable manifold of saddle-points, but they have their limitations, especially when the system is not invertible. In this paper, we present a new algorithm to compute the stable manifold of 2-dimensional systems which can also be used for non-invertible systems. After outlining the logic of the algorithm, we demonstrate the output of the algorithm on several examples.\\

	\end{abstract}	

	\vspace{5mm}
	
\begin{multicols*}{2}
\section{Introduction}
 
 \noindent Saddle fixed points play a vital role in determining the behaviour of
dynamical systems. These are points that act as attractors in some
directions and repel trajectories in other directions. This
information is conveyed in the stable and the unstable manifolds of
the saddle point. For a 2-D map, an unstable manifold is a one-dimensional subset with the property that (a) it is tangential to the unstable eigenvector of
the concerned fixed point, (b) iterates starting from any point on
that manifold always remain on the manifold, and (c) iterates
progressively move away from the fixed point along this
manifold. Similarly, the stable manifold is defined as a subset with
the property that (a) it is tangential to the stable eigenvector of
the concerned fixed point, (b) iterates starting from any point on
that manifold always remain on the manifold, and (c) iterates
progressively move closer to the fixed point along this manifold. 

It is known that an unstable manifold has attractive property and all
attractors in a system occur on an unstable manifold of a fixed
point. A stable manifold, on the other hand, has repelling property
and if a system has multiple attractors, a stable manifold forms the
boundary between the basins of attraction of the attractors. The
interplay between the stable and unstable manifolds is responsible for
important dynamical phenomena like interior crises, boundary crises,
homoclinic intersections, etc.  Hence, computing the structure of the
stable and unstable manifolds of saddle points assumes special
importance in understanding the dynamics of systems.

It is generally not possible to obtain the one-dimensional manifolds
(higher dimensional manifolds are not considered here) of saddle fixed
points analytically, and hence, must be computed numerically. It is
simpler to obtain the unstable manifold: The technique generally starts with
approximating an initial segment $s$ of the manifold close to the
saddle point with the aid of the eigenvector corresponding to the
unstable eigenvalue of the local linearization of the system at the
point. Different algorithms can then be applied to extend the manifold
from this initial segment. Over the years, this ``manifold extension
algorithm'' has seen a lot of variation, as techniques were developed
to obtain sufficiently closely spaced points so that the manifold can
be sketched \cite{dynamics-manual}.

It may be noticed that the definition of the stable manifold becomes
identical with that of the unstable manifold under the application of
the inverse map. That is why the same method can be applied to compute
the stable manifold by iterating backward using the inverse map. This
is the method used in the software tool \textit{Dynamics}
\cite{dynamics-manual} to sketch the stable manifold. But that
requires the map to be invertible, which may not always be the
case. To overcome the problem, a few alternative techniques have been
proposed.

One of the proposed algorithms \cite{kostelich1996} computes the stable manifold through the use of local inverse(s) for computing a sequence of pre-images of points on the segments. Thus it avoids the use of a global definition of the
inverse map. However, this algorithm cannot be applied in cases
where local inverses cannot be found.

In an alternative method, the
initial segment is extended by using a ``prediction and correction''
technique \cite{li-2012} that uses the inverse of the local linearization of
the given system to first predict a preimage of the current point, and
then `correct it' so that it is close to the actual preimage. However,
as this technique also uses the inverse system, it faces similar
limitations as \cite{kostelich1996}. Alternatively, the algorithm described in
\cite{lien1998} may be used to compute the pre-images of points (even in case
of non-invertible maps) on the segment $s$ to obtain a set of points
that lie on the stable manifold.

The Search Circle Algorithm \cite{search-circle} extends the stable
set, starting with $s$, by finding points at a particular distance
from the latest point in the set that map back to a region already
included in the stable set. The Search Circle algorithm was
implemented on a software tool called DSTool \cite{dstool}. It is a
powerful method that works well for most maps --- invertible or
non-invertible. But its performance may be limited in case there are
sharp bends and folds in the structure of the stable
manifold. Moreover, it is no longer convenient to use DSTool as it
works only on 32-bit architecture.

Thus we see that, even though some algorithms have been developed to
compute the stable manifolds of non-invertible maps, these have some
limitations because of the requirement of computing the pre-iterates in
roundabout ways, necessitating complicated implementations that need
strong error corrections to maintain accuracy. Even then, most of
these algorithms fail when a stable manifold has sharp bends and
folds.

In this paper, we present a simple and effective algorithm that uses
only forward iterates to compute the stable manifold and thus is free
from the problems of computation of pre-iterates of non-invertible
maps. We call it the `Point-Iterative Algorithm', which uses the
properties of the stable manifold to reduce the problem at hand to a
computationally simpler problem. We demonstrate the performance of the
algorithm on different systems and show that the errors occurring
through this algorithm are minimal.

\section{The Proposed Logic}
For any study of a dynamical system, it is seldom necessary to plot
the structure of the stable and unstable manifolds over the
entire phase plane. Normally the necessity is to plot the manifolds in
smaller regions of interest. Once this is achieved, a more complete
sense of the dynamics of the system can be obtained by consolidating
the behaviour observed in all these smaller regions. Hence, for all practical
purposes, we need to sketch the stable manifold in a smaller region of
the phase plane such that,
\begin{gather}
a_1\le x\le a_2\\
b_1 \le y \le b_2,
\end{gather}
where $x$ and $y$ are the state variables and the specification of the
system (whether in the form of differential equations or maps)
provides the rules for the evolution of these state variables.

In our algorithm, we first compute the points where the
stable manifold intersects the boundary of this region. 
Then the system can be simply evolved forward starting from these
points of intersection, and the plot so obtained would be the sketch
of the stable manifold in that region.

So, the problem is now reduced to that of finding the points of
intersection of the manifold with the boundary of a given region. For
this, we use the fact that the stable manifold acts as a repeller of
trajectories, i.e., trajectories starting at very small distances on
opposite sides of the stable manifold end up with very different
fates. Using this property, the points of intersection are found,
which allows us to compute the stable manifold using only the forward
iterates.

\subsection{The Parameters}\label{Parameters}

\begin{enumerate}
\item 
BISECTION\_ERROR: This describes the maximum error that is allowed in the bisection. Usually, 1e-6 is a good value for the bisection error.
\item
X\_LINEAR\_ADVANCEMENT: This describes the step length along the horizontal axis. Usually, 3-7\% of the length of the horizontal boundary is a good measure to start with.
\item
Y\_LINEAR\_ADVANCEMENT: This describes the step length along the vertical axis. Usually, 3-7\% of the length of the vertical boundary is a good measure to start with.
\item
N\_MAX: The utility of this parameter is described later. Usually, N\_MAX = 5 or 6 is a good starting point.
\item
points[ ]: This is the set of the points of intersection of the stable manifold with the boundary of the region.    
\end{enumerate}

\subsection{The Functions}\label{Functions}
\begin{enumerate}
\item 
evolve\_system (initial, system): Given the system definition and the initial conditions, it calculates the N\_MAX-th iterate of the given map. It returns the vector pointing towards the next point in the system. 

\item
check\_for\_opposite\_sides (p1,p2, system): Given two points, p1 and p2 and the system description, this function checks whether p1 and p2 are on opposite sides of the stable manifold. It does so by evolving p1 and p2 under evolve\_system, and checking whether the vectors corresponding to the two points are diverging at least along one of the horizontal and vertical directions. F
Sometimes, it may so happen that due to the particular choice of N\_MAX, such points may be obtained, which are not actually on opposite sides of the stable manifold but the vectors happen to diverge for the current parameter values. It might be a good idea to check for two or more N\_MAX values before applying the bisection procedure.

\item
bisect (p1,p2,mode,system): Given two points, p1 and p2, it performs bisection between p1 and p2 and adds all points of intersections to the set points[ ]. The mode decides whether the bisection is performed along the horizontal or the vertical axis. 

\item
find\_points\_of\_intersection (p1,p2,mode,system): Given the endpoints of the region boundary, this uses bisect to find all the points of intersection along with the boundary. The different values of mode correspond to the two directions of the boundary.

\begin{center}
    \begin{tabular}{|c|c|}
    \hline
        Mode & Direction\\
    \hline
        0 & Horizontal Axis\\
        1 & Vertical Axis\\
        All other values & Invalid Mode\\
    \hline
    \end{tabular}
\end{center}
\end{enumerate}

\noindent { The pseudocodes of all these functions are given in the Appendix.}

\subsection{Special Consideration for Maps}

While the proposed logic of finding the points of intersection and
evolving the system can be applied to differential equations as well
as maps, some special considerations are necessary when it is applied
to maps.

In case of a continuous time system,
  theoretically an infinite number of points lying on the stable
  manifold can be found through evolution of a point on the
  manifold. In order to sketch the manifold computationally, a finite
  set of discrete points is obtained using a numerical algorithm,
  which is a subset of the infinite number of points. The number of
  points in the set can be increased to any desirable extent so as to
  obtain an accurate plot of stable manifold, i.e., the dicretization
  is under the control of the programmer.

  In case of maps, that is not the case.  Due to the discrete nature
  of maps, iteration of the points of intersection of the stable
  manifold with the boundary of the region of interest results in a
  finite number of points lying on the stable manifold. The result is a
  discrete scatter of points from which it might be difficult to
  infer a pattern.

For this reason, if a reasonably accurate and (almost) continuous plot
is to be obtained, we need to run the
\textit{find\_points\_of\_intersection} for all distinct regions of
size X\_LINEAR\_ADVANCEMENT $\times$ Y\_LINEAR\_ADVANCEMENT, present
within our region of interest. Hence, the
\textit{sketch\_stable\_manifold} function takes a slightly different
form in case of maps.  
sketch\_stable\_manifold function for maps is provided in
This algorithm is a simple, yet powerful
technique that can sketch complicated structures of the stable
manifold, some of which are demonstrated in Section \ref{Examples}.

\section{Examples}\label{Examples}

We now demonstrate the performance of the Point-Iterative algorithm with the plots obtained for 2-dimensional maps. We verify the plots by sketching the basins of attraction alongside (since the stable manifold lies along the basin boundaries), and comparing their shapes.

\begin{figure*}[]
    \centering
    \subfigure[]{\includegraphics[width=6.4cm, height=6.4cm]{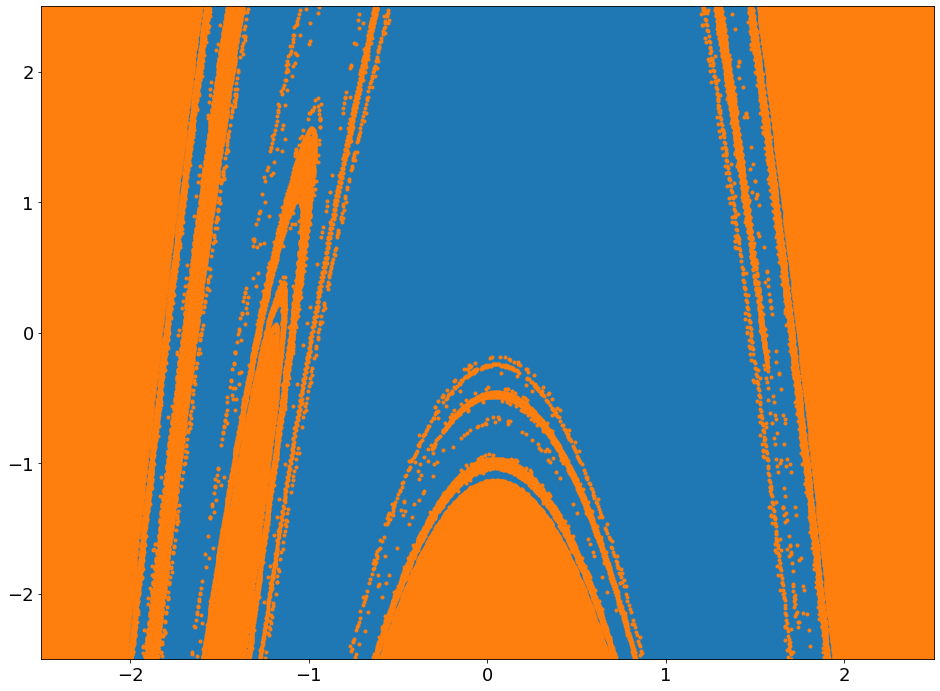}}
    \subfigure[]{\includegraphics[width=6.4cm, height=6.4cm]{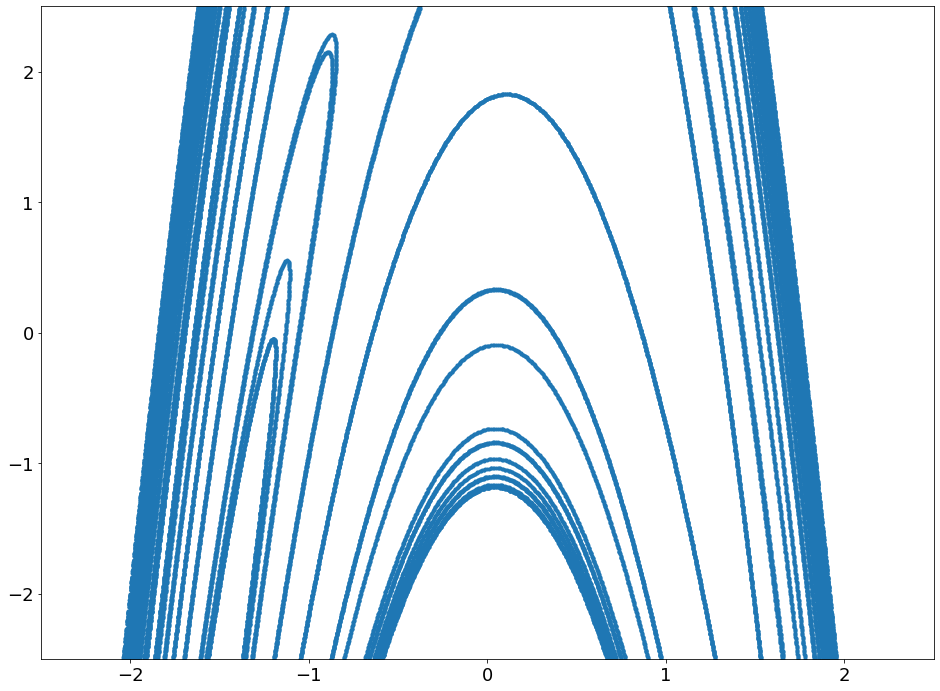}}
    \subfigure[]{\includegraphics[width=6.4cm, height=6.4cm]{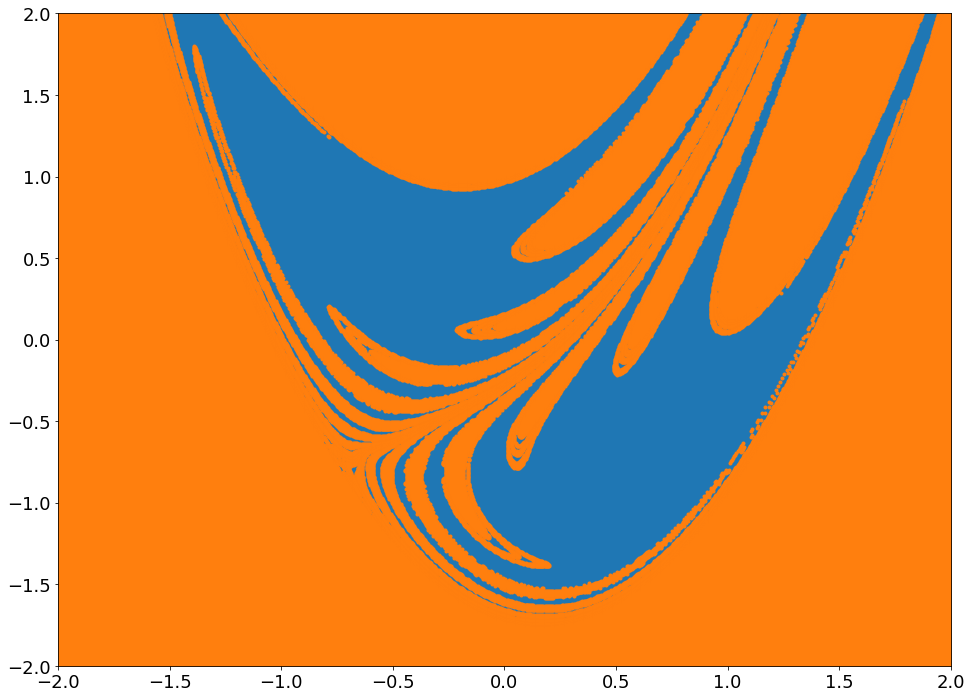}}
    \subfigure[]{\includegraphics[width=6.4cm, height=6.4cm]{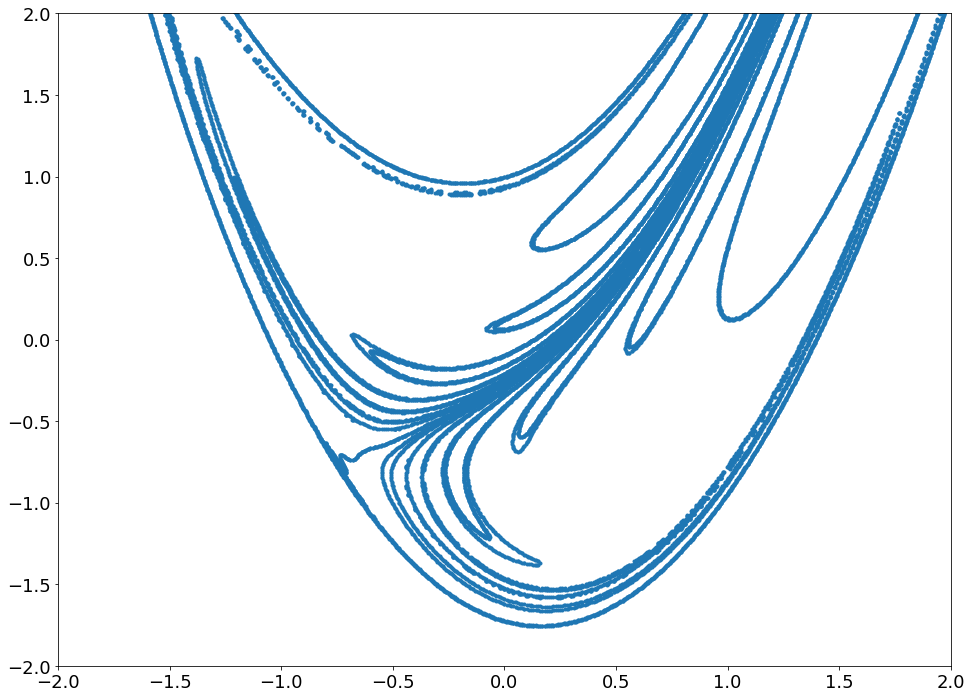}}
    \subfigure[]{\includegraphics[width=6.4cm, height=6.4cm]{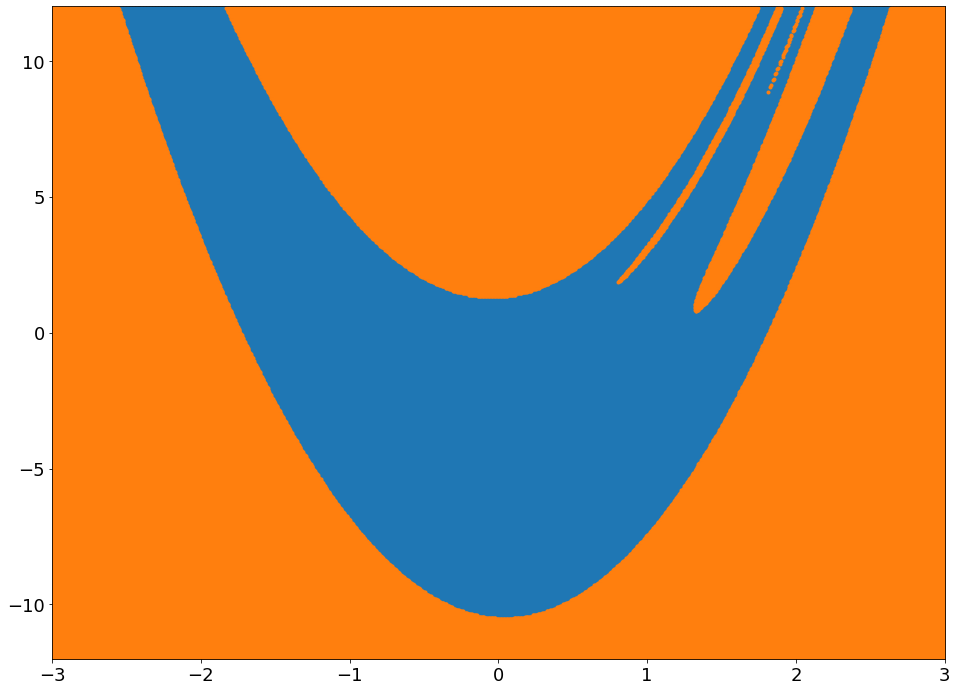}}
    \subfigure[]{\includegraphics[width=6.4cm, height=6.4cm]{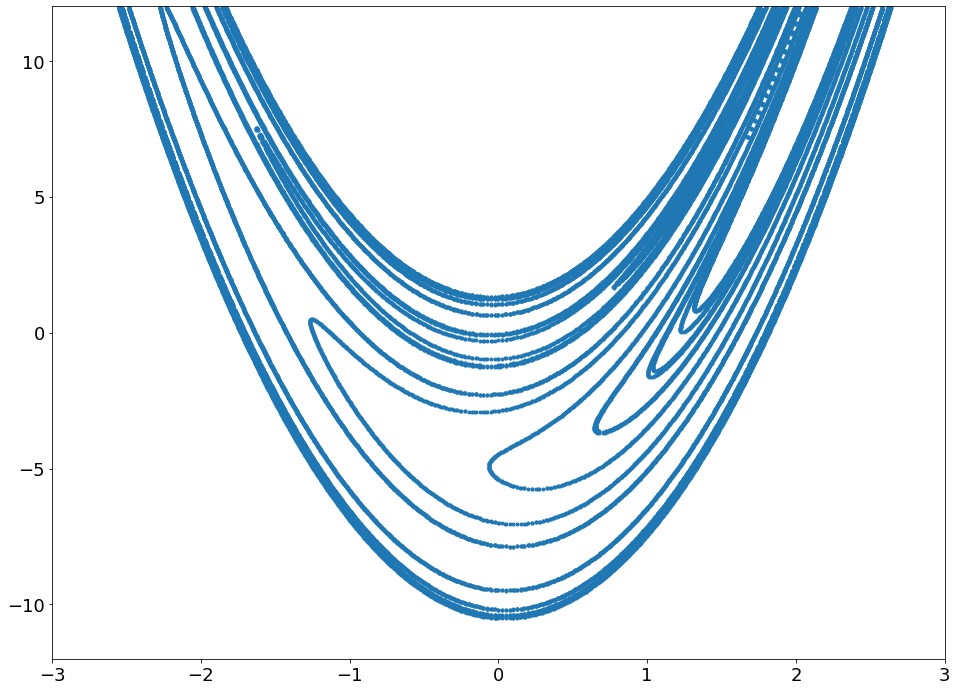}}
    \caption{(a) and (b): The basins of attraction and the stable manifold plotted by the Point-Iterative Algorithm for the Henon Map for $a=1.4$, $b=-0.3$.
    (c) and (d): The same for parameter values $a=0.5$, $b=0.9$. (e) and (f): The same for parameter values $a=1.42$, $b=0.3$.}
    \label{Henon-figs}
\end{figure*}

\subsection{Henon Map}
\begin{figure*}[]
    \centering
    \includegraphics[width=6.4cm]{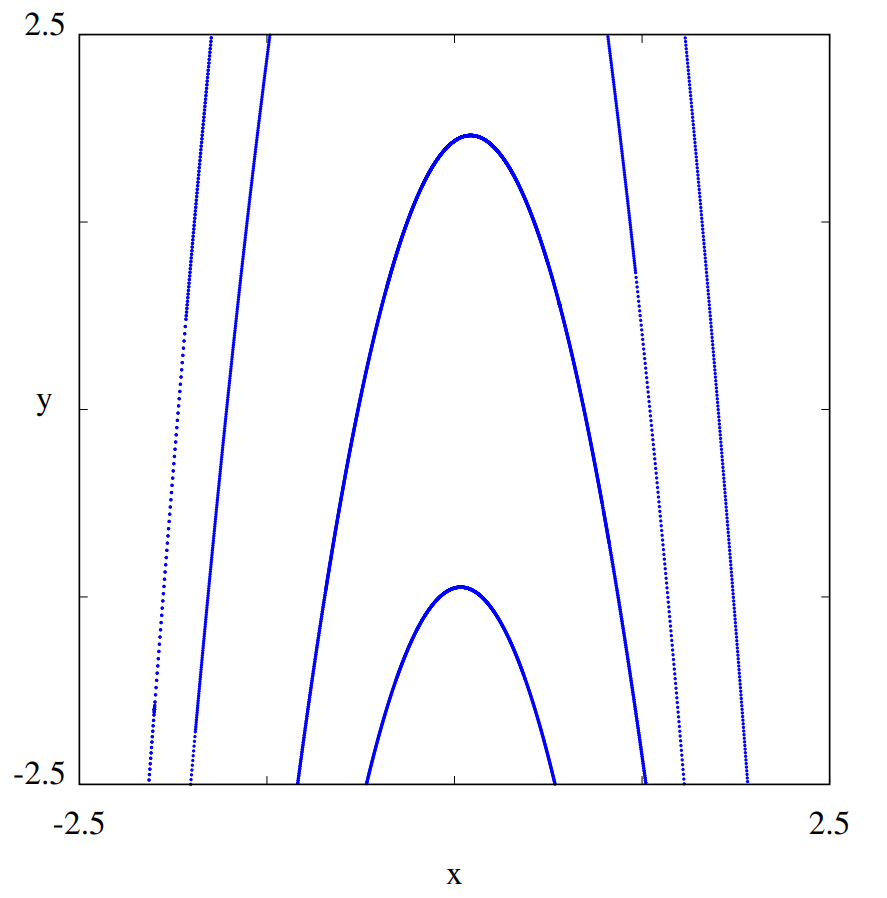}
    \caption{Stable manifold of the Henon Map for $a=1.4$ and $b=-0.3$ obtained using the Search-Circle Algorithm}
    \label{Search-Circle-Plot}
\end{figure*}

For the parameter values we have considered, the Henon map is invertible  but has relatively sharp bends and folds. The map is described by
\begin{align}
    \begin{pmatrix}
    x_{n+1}\\
    y_{n+1}
    \end{pmatrix}
    &=
    \begin{pmatrix}
     a-x_n^2+by_n\\
     x_n        
    \end{pmatrix},
    \label{henon}
\end{align}
where $a$ and $b$ are parameters which control the behaviour of the
system. In Fig.~\ref{Henon-figs}, we have plotted the basins of attraction and the stable manifolds obtained by our algorithm for three different
values of $(a,b)$. The obtained manifolds are found to lie on the basin boundaries, and additionally show intricate structures that are not visible in the basin plots because of the finite resolution of the basin images. The structure of the stable manifold for $a=1.4$ and $b=-0.3$ is also compared with the manifold obtained from DSTool \cite{dstool} using the Search-Circle Algorithm \cite{search-circle} (Fig.~\ref{Search-Circle-Plot}). As can be seen from Fig.~\ref{Henon-figs}(b), the sharp bends and folds of the manifold are missing in Fig.~\ref{Search-Circle-Plot}.

\subsection{Modified Ikeda Map}
The Ikeda map is described by:
\begin{align}
    \begin{pmatrix}
    x_{n+1}\\
    y_{n+1}
    \end{pmatrix}
    &=
    \begin{pmatrix}
     a+bx_n\cos(m)-ey_n\sin(m)\\
     by_n\cos(m) + ex_n\sin(m)        
    \end{pmatrix}
\end{align}
where 
\begin{equation}
m = \phi - \frac{q}{1+x_n^2+y_n^2}    
\end{equation}
and $a,b,e,q,\phi$ are parameters that control the behaviour of the map. 
For the typical Ikeda map, $b$=$e$, and as mentioned in \cite{search-circle}, it is possible to find an explicit inverse. However, we consider slightly different values for $b$ and $e$ for which no closed analytic form of the inverse exists \cite{search-circle,kostelich1996}.

\begin{figure*}[]
    \centering
    \includegraphics[width=7cm, height=7cm]{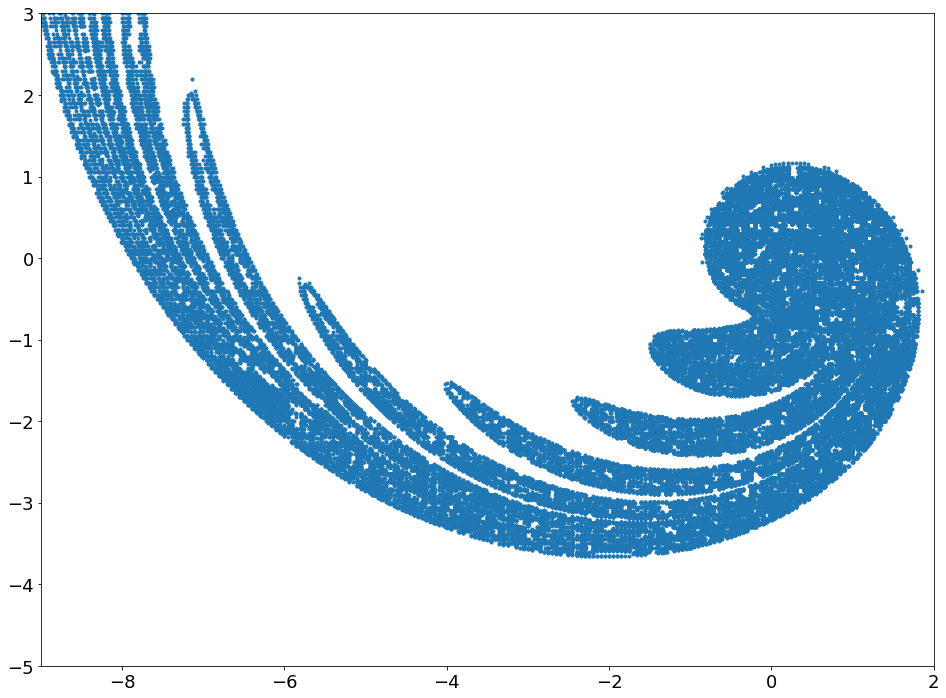}
    \caption{Sketch of the stable manifold for the modified Ikeda map for $a=1$, $b=0.9$, $\phi=0.4$, $q=6$ and $e=1$, obtained using the Point-Iterative algorithm.}
    \label{Ikeda-Manifold}
\end{figure*}

Considering the values $a=1$, $b=0.9$, $\phi=0.4$, $q=6$ and $e=1$, we
have obtained a plot of the stable manifold using the Point-Iterative
Algorithm (Fig.~\ref{Ikeda-Manifold}). This is very similar to the
plot in \cite{kostelich1996} and in \cite{search-circle} (generated by the Search Circle Algorithm).

\subsection{Modified Gumowski-Mira Map}\label{gumowski-mira}
The Gumowski-Mira map is expressed as:

\begin{align}
    \begin{pmatrix}
    x_{n+1}\\
    y_{n+1}
    \end{pmatrix}
    &=
    \begin{pmatrix}
     y_n\\
     ax_n + bx_n^2 + y_n^2        
    \end{pmatrix}
\end{align}
where $a$ and $b$ are parameters which control the behaviour of the system.
It is easy to notice that while finding the inverse of the above map, we obtain a quadratic equation of $x_n$ in terms of $x_{n+1}$ and $y_{n+1}$. Hence, the map is non-invertible as there might exist two distinct pre-images of a single point. Algorithms that make use of the inverse for sketching the stable manifold fail to work here.

\begin{figure*}[]
    \centering
    \includegraphics[width=7.5cm,height=7.5cm]{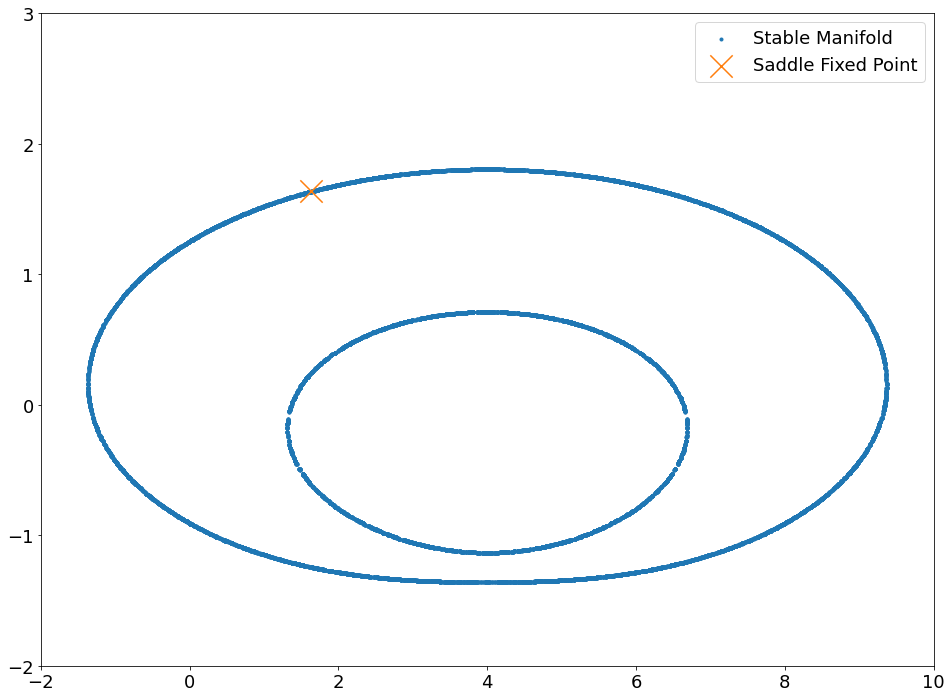}
    \caption{Sketch of the stable manifold of the saddle-fixed point (marked in orange) of the modified Gumowski-Mira Map for $a$=-0.8, $b$=0.1 obtained using the Point-Iterative Algorithm}
    \label{Gumowski-Mira}
\end{figure*}

Using $a=-0.8$ and $b=0.1$, we observe that there exists a saddle-fixed point at $(x,y) = (1.636, 1.636)$. Fig.~\ref{Gumowski-Mira} shows the sketch of the stable manifold of this saddle point generated by the Point-Iterative Algorithm. The manifold has disjoint sets which is a result of the existence of two pre-images. Interestingly, the branches of the stable manifold on both sides of the saddle-point join to form a closed loop. The stable manifold plotted by our algorithm has a good correspondence to the plot in \cite{search-circle} generated by the Search-Circle Algorithm.

\begin{figure*}[t]
    \centering
    \subfigure[]{\includegraphics[width=7.5cm, height = 7.55cm]{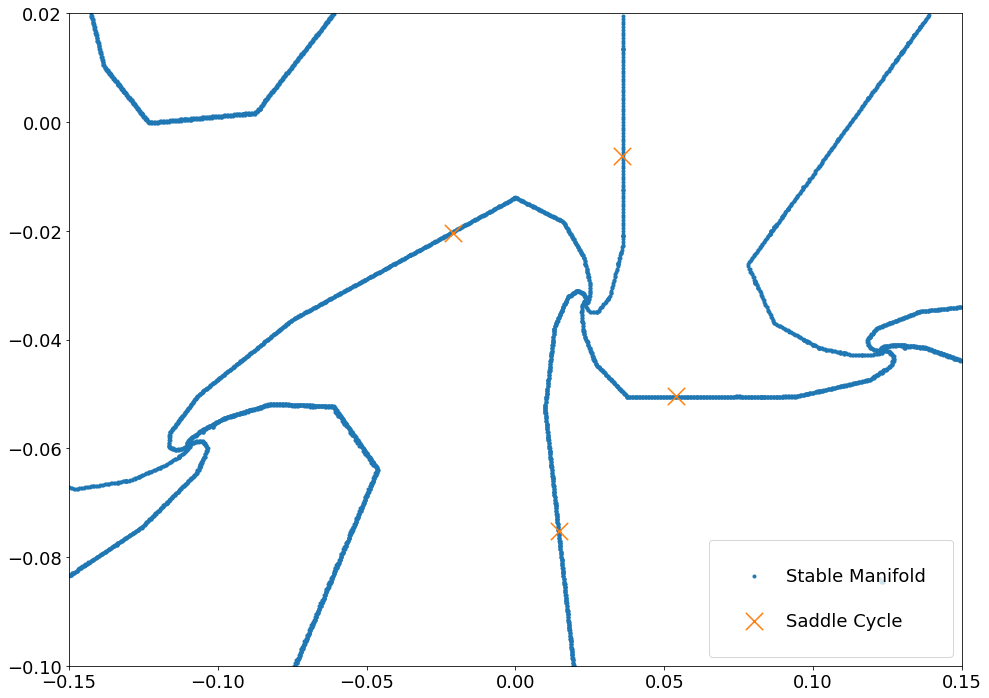}}
    \subfigure[]{\includegraphics[width=7.5cm, height = 7.5cm]{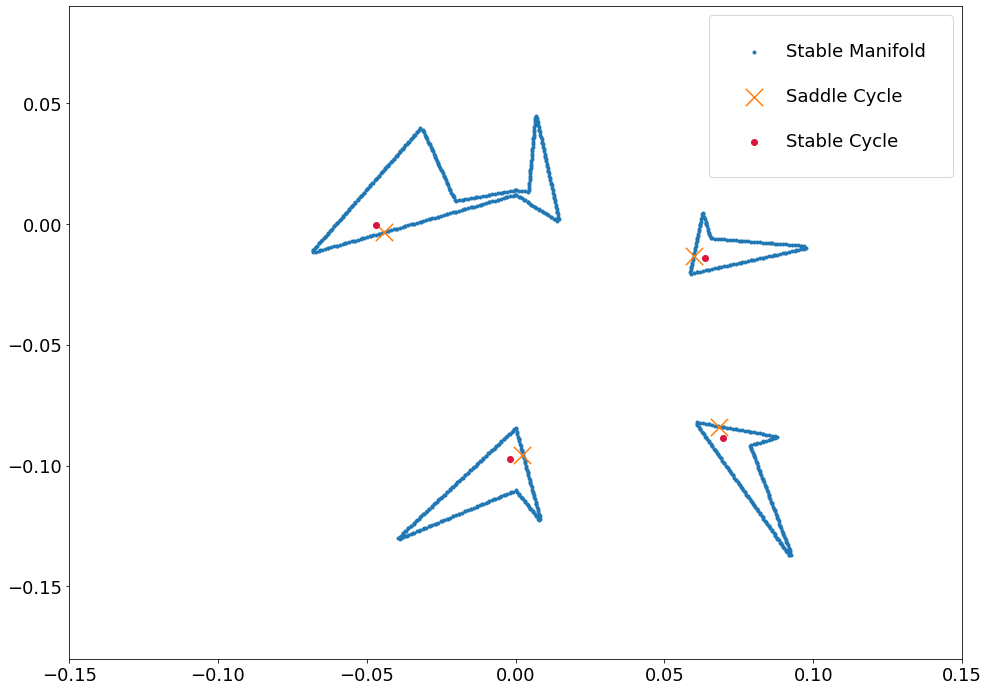}}
    \caption{Sketch of the stable manifold for the non-smooth map for parameters $\delta_L=-0.3,\ \tau_L=-0.3,\ \delta_R=1.4,\ \mu=0.05$ and (a) $\ \tau_R=0.28$, (b) $\ \tau_R=0.53$}
    \label{Non-Smooth-Map-Manifold}
\end{figure*}

\subsection{Non-Smooth non-invertible Map}
Non-smooth maps have been subject of intense investigation in recent times because of their applicability in many physical and engineering systems. The piecewise linear `normal form' of the non-smooth map is given by
\begin{align}
    \begin{pmatrix}
    x_{n+1}\\
    y_{n+1}
    \end{pmatrix}
    &=
    \begin{cases}
     h_1(x_n,y_n) & if x < 0\\
     h_2(x_n,y_n) & if x\geq 0
    \end{cases}
\end{align}
where
\begin{align*}
    h_1(x_n,y_n)
    &=
    \begin{pmatrix}
     \tau_Lx_n + y_n + \mu\\
     -\delta_Lx_n
    \end{pmatrix},\\
    h_2(x_n,y_n)
    &=
    \begin{pmatrix}
     \tau_Rx_n + y_n + \mu\\
     -\delta_Rx_n
    \end{pmatrix}
\end{align*}
and $\delta_L,\ \tau_L,\ \delta_R,\ \tau_R$ and $\mu$ are parameters
that determine the behaviour of the map. The map becomes
non-invertible if $\delta_L$ and $\delta_R$ are of opposite sign or if
one of them is zero \cite{de2012}.  We take the parameter values
$\delta_L = -0.3,\ \tau_L = -0.3,\ \delta_R = 1.4,\ \tau_R = 0.28$ and
$\mu = 0.05$, satisfying this condition. The phase portrait of the
system is characterised by saddle-node connections of a pair of
period-4 cycles (a stable cycle and a saddle
cycle). Fig.~\ref{Non-Smooth-Map-Manifold}(a) shows the plot of the
stable manifold of the non-smooth map for the above parameter values
as generated by the Point-Iterative Algorithm. The plot largely
resembles the sketch of the stable manifold in \cite{de2012} that were
obtained using DSTool. Our algorithm could detect some additional
sections (compare Fig.~\ref{Non-Smooth-Map-Manifold}(a) with Fig.~7(a) of \cite{de2012}) which have been verified to belong to the stable
manifold.

For parameter values $\delta_L = -0.3,\ \tau_L = -0.3,\ \delta_R =
1.4,\ \tau_R = 0.53$ and $\mu = 0.05$, the phase portrait consists of
a chaotic attractor and a stable period-4 cycle with their basins of
attraction separated by the stable manifold of a saddle period-4
cycle. Fig.~\ref{Non-Smooth-Map-Manifold}(b) shows the stable and saddle period-4 cycles and the stable manifold obtained through the Point-Iterative
Algorithm. More details about the `closed loop' shape of the stable manifold can be found in \cite{de2012}. The plot has an exact correspondence to the basins of attraction shown in
\cite{de2012}. It must be noted here that the logic for the \textit{check\_for\_opposite\_sides} function used in the Appendix
will produce several false-positives if
there is a chaotic attractor present in the phase portrait. Hence, to
obtain the plot in Fig.~\ref{Non-Smooth-Map-Manifold}(b), the
implementation of
\textit{$check\_for\_opposite\_sides(p_0,p_1,system)$} was modified to
return {\em True} if trajectories starting from $p_0$ were attracted to
the stable cycle and those starting from $p_1$ were not, and
vice-versa. The function would return {\em False} otherwise.

\subsection{Map obtained from a 3 dimensional continuous time system}

In order to demonstrate the power of this algorithm, we use it to plot the stable manifold of the map obtained by placing a Poincare section on the phase space of the continuous time system which is expressed as:
\begin{align}
    \begin{pmatrix}
    \dot x\\
    \dot y\\
    \dot z
    \end{pmatrix}
    &=
    \begin{cases}
    y\\
    z\\
    -y + 0.1x^2 + 1.1xz + 1.05
    \end{cases}
\end{align}

\begin{figure*}[]
    \centering
    \subfigure[ ]{\includegraphics[width=5cm]{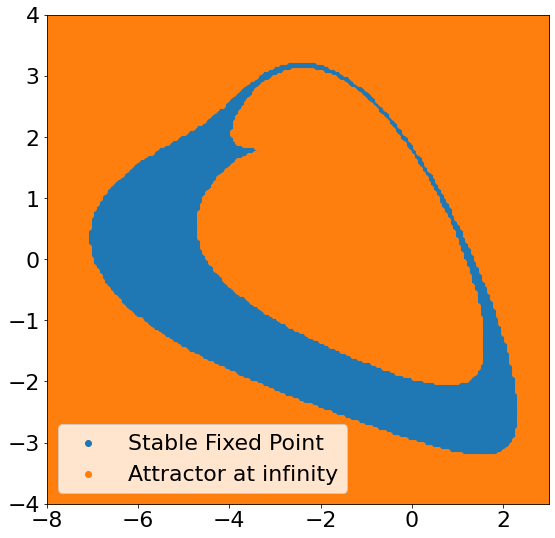}}\hspace{0.5in}
    \subfigure[ ]{\includegraphics[width=5cm]{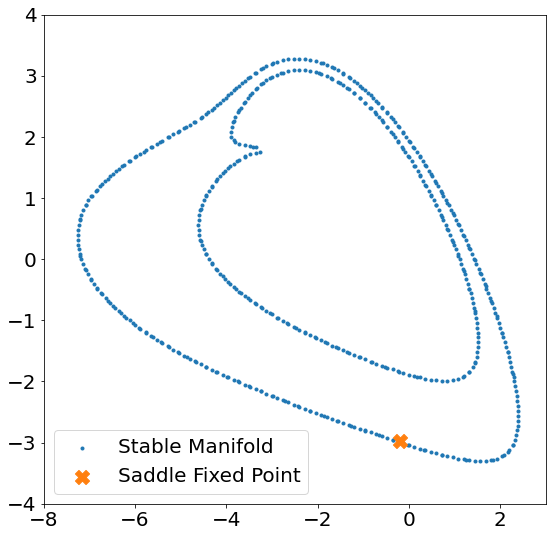}}
    \caption{(a) Basins of attraction and (b) Stable Manifold of the saddle fixed point of the planar map obtained from the 3D continuous time system}
    \label{Continuous-Manifold}
\end{figure*}

This continuous time system has attracted some research attention because it does not have any equilibrium point and yet has a stable and an unstable limit cycle and an attractor at infinity. Upon placing a Poincar\'e section $z=-2$ on the phase plane, we obtain a planar map with two fixed points. The map has a stable fixed point at $(0.2459,-2.4506)$ and a saddle fixed point at $(-0.2036, -2.93)$, the stable manifold of which separates the basins of attraction of the stable fixed point and the attractor at infinity. The basins of attraction and the stable manifold of the saddle fixed point obtained using the Point-Iterative Algorithm have been shown in Fig.~\ref{Continuous-Manifold} and the stable manifold is found to lie along the basin boundaries. This demonstrates the usefulness of the algorithm as it does not need a closed form expression of the 2-dimensional map to obtain the stable manifold.

\section{Adjusting the parameters used by the Algorithm}\label{Parameter Adjustment}
Section \ref{Parameters} lists the parameters that are used by the algorithm to sketch the stable manifold. For some systems, the initial setting of parameters might produce the desired results. However, for others, some tuning may be necessary. We now list how changing the various parameters affects the plot obtained from the algorithm.

\begin{figure*}[tbh]
    \centering
    \subfigure[]{\includegraphics[width=5.8cm, height=5.2cm]{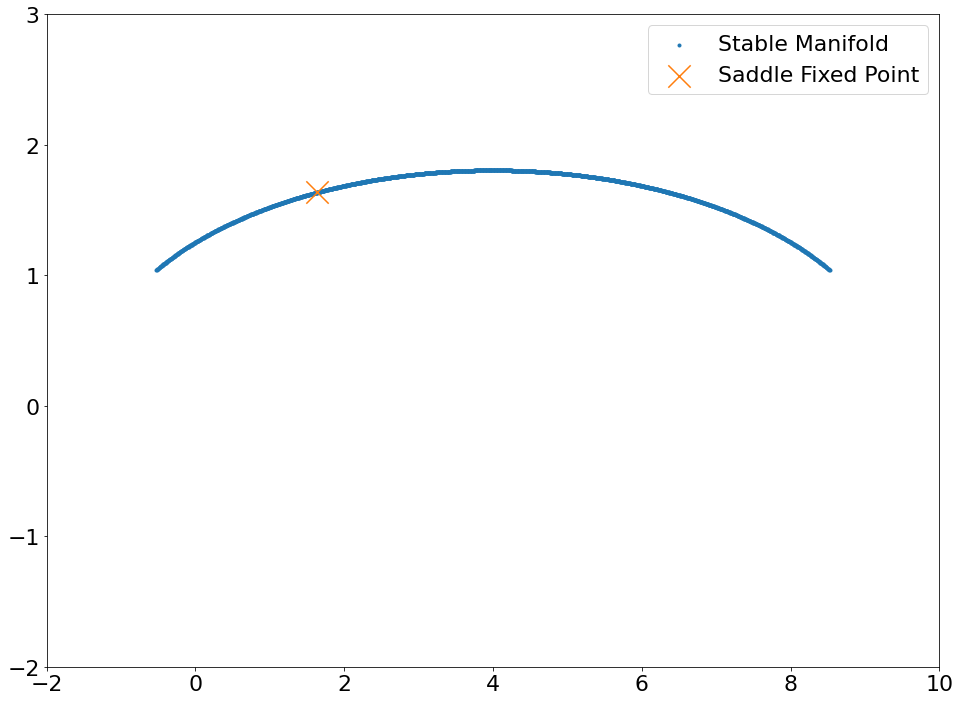}}
    \subfigure[]{\includegraphics[width=5.8cm, height=5.2cm]{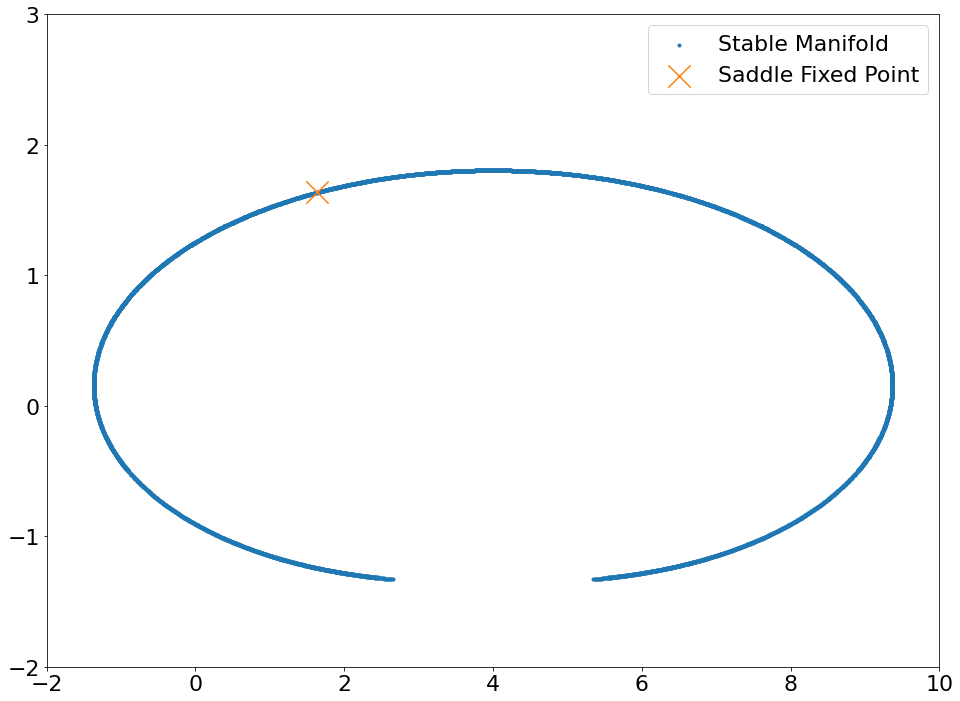}}  
    \subfigure[]{\includegraphics[width=5.8cm, height=5.2cm]{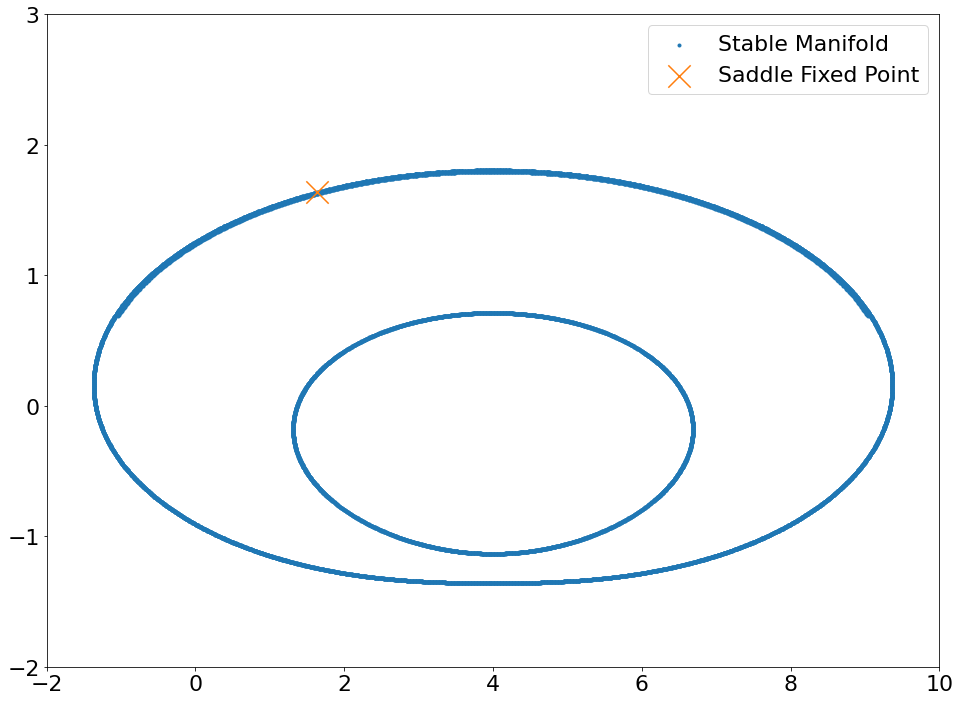}}
    \subfigure[]{\includegraphics[width=5.8cm, height=5.2cm]{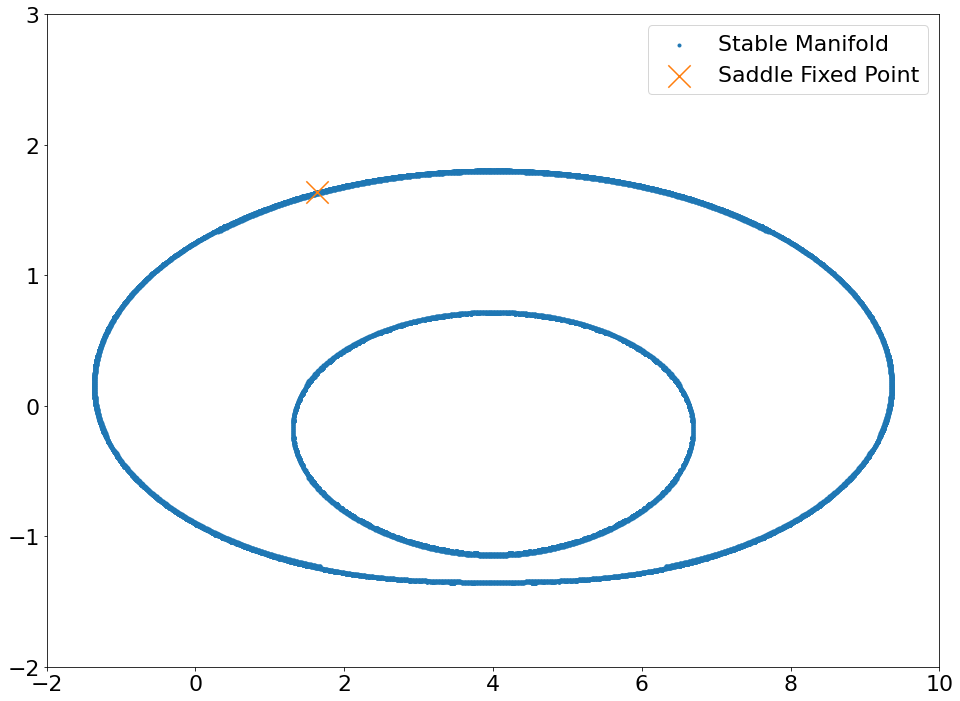}}
    \caption{Plot of the stable manifold obtained using the algorithm for (a) $N\_MAX$ = 3 (b) $N\_MAX$ = 4 (c) $N\_MAX$ = 5 (d) $N\_MAX$ = 14. 
    }
    \label{Choosing-nmax}
\end{figure*}

\begin{enumerate}
    \item 
      BISECTION\_ERROR: As this parameter controls the accuracy to
      which a point of intersection is bisected, very small values
      ($<$\:1e-7) might increase the computation time, and thereby,
      affect the performance of the algorithm. Whereas a larger value
      for this parameter might mean a larger error in the position of
      the point of intersection. Usually, values
        between 1e-4 and 1e-6 achieve an acceptable compromise between
        computation time and accuracy.

    \item
    X and Y LINEAR\_ADVANCEMENTS: These essentially control the size of the steps that are taken along the boundary, before carrying out bisection. While reducing their values might lead to a greater computation time, for large values of X and Y LINEAR\_ADVANCEMENTS ($>$15-20\% of the boundary length), the stable manifold for maps may be reduced to a discrete sequence of points. These large values may be used when it is required to find out the general outline of the stable manifold. However, if the actual shape manifold needs to be plotted, their values need to be reduced.
    
    \item
    N\_MAX: This is essentially used to identify whether two points
    are on opposite sides of the stable manifold. If trajectories
    starting at these two points are found to diverge along at least
    one of the directions, after N\_MAX iterates, the points are said
    to be on opposite sides of the stable manifold. Hence, a small
    value ($<$3) might see the stable manifold breaking off
    unexpectedly in the phase plane, i.e., a discontinuous manifold
    may be obtained. On the other hand, a very large value would
    demand a large computation time. We illustrate the
      issue with reference to the modified Gumowski-Mira map discussed
      in Sec.~\ref{Examples}. Fig.~\ref{Choosing-nmax} shows the plots
      of the stable manifold of the above map for the parameter values
      specified in Sec.~\ref{gumowski-mira} obtained using the
      Point-Iterative Algorithm for different values of $N\_MAX$. As
      can be seen, for $N\_MAX$ = 3 and $N\_MAX$ = 4, the plots
      obtained break off in the middle, indicating that the value of
      $N\_MAX$ must be increased. However, at $N\_MAX$ = 5, the
      manifold does not break off in the middle and we obtain the
      structure of the stable manifold. In
      Fig.~\ref{Choosing-nmax}(d), the stable manifold is computed for
      a higher value of $N\_MAX$ (=14). As this shows no additional
      structure in the manifold, the chosen value of $N\_MAX$ is 5 as
      this is the smallest value of $N\_MAX$ that computes the
      structure of the stable manifold in the minimum computation
      time. Hence, the value of $N\_MAX$ should be so chosen that it
      results in a continuous manifold for a continuous map, and a
      further increase in its value does not provide any additional
      information about the stable manifold. While $N\_MAX$ = 5 or 6 may
    be a good starting point, some maps may require an $N\_MAX$ value
    up to 15 or even in some cases, as high as 25-30.
\end{enumerate}

It must be noted that the exact values of these parameters may vary from system to system. However, the general behaviour upon increasing/decreasing the values remains similar. Hence, these should be used as indicators for tuning the parameter values until the required plot is obtained.

\subsection{Accuracy of the Plots Obtained}
While it is necessary to estimate and through the use of necessary techniques, confine the errors within certain bounds with the use of certain approximations, it is also necessary to ensure that heavy computational resources are not lost in overcoming these errors.

The use of various approximate algorithms such as approximating the local inverse \cite{kostelich1996} or the Search Circle \cite{search-circle} to extend a preliminary section of the stable manifold are followed by strong error analysis to ensure that the plots obtained are accurate. However, the Point-Iterative algorithm does not need rigorous error analysis. We find the points of intersection at multiple steps, and the only error encountered here is the uncertainty in the position of the point, i.e., the BISECTION\_ERROR. Since the BISECTION\_ERROR usually lies between 1e-4 and 1e-6, the plots obtained are, for all practical purposes, quite accurate.

\begin{table*}[b]
    \centering
    \begin{tabular}{|l|}
        \hline
        \includegraphics[width=3.5in]{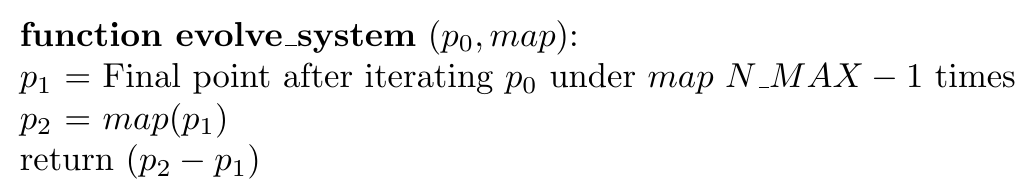} \\
        \hline
        \includegraphics[width=3.5in]{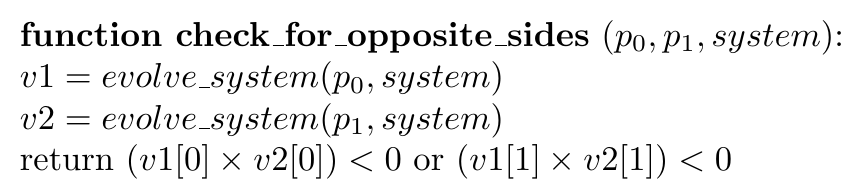} \\
        \hline
        \includegraphics[width=3.5in]{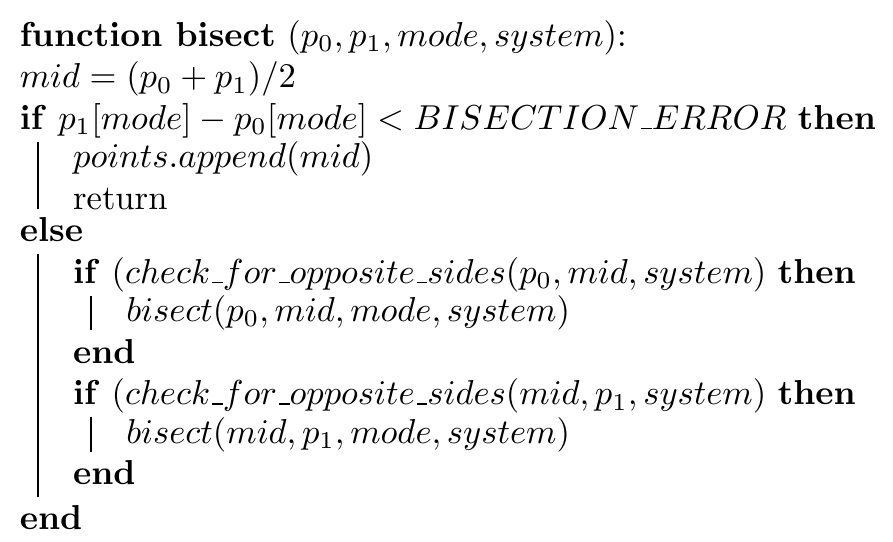} \\
        \hline
        \includegraphics[width=3.5in]{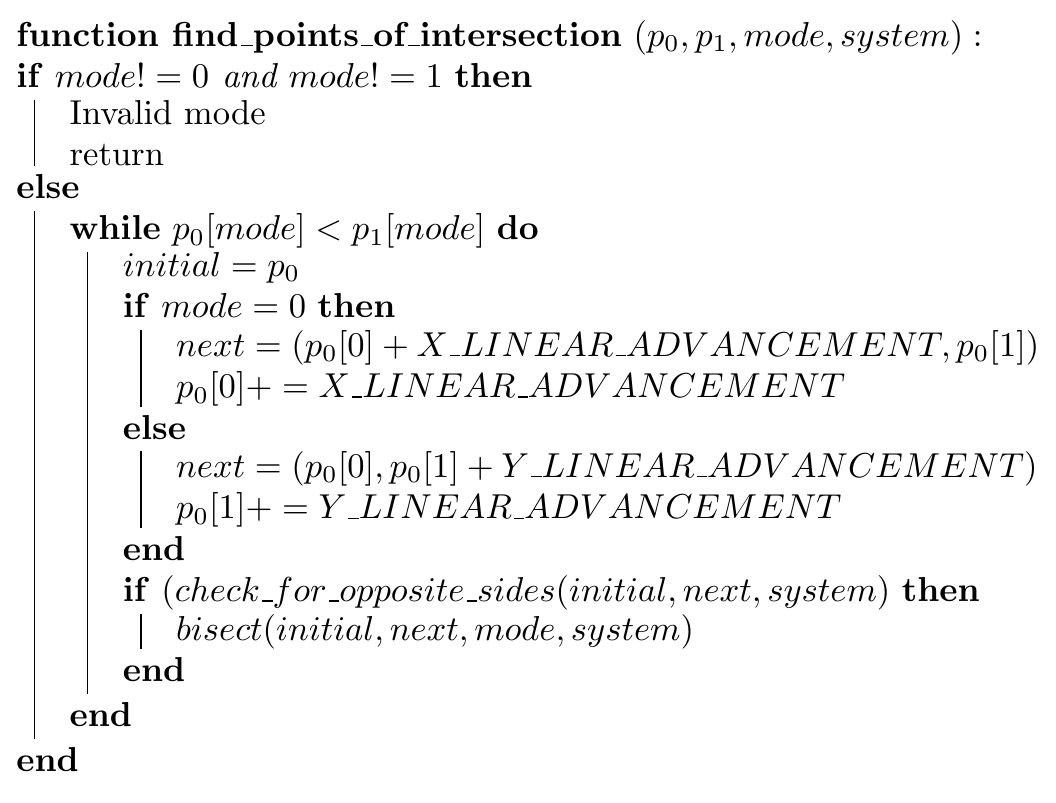} \\
        \hline
        \includegraphics[width=3.5in]{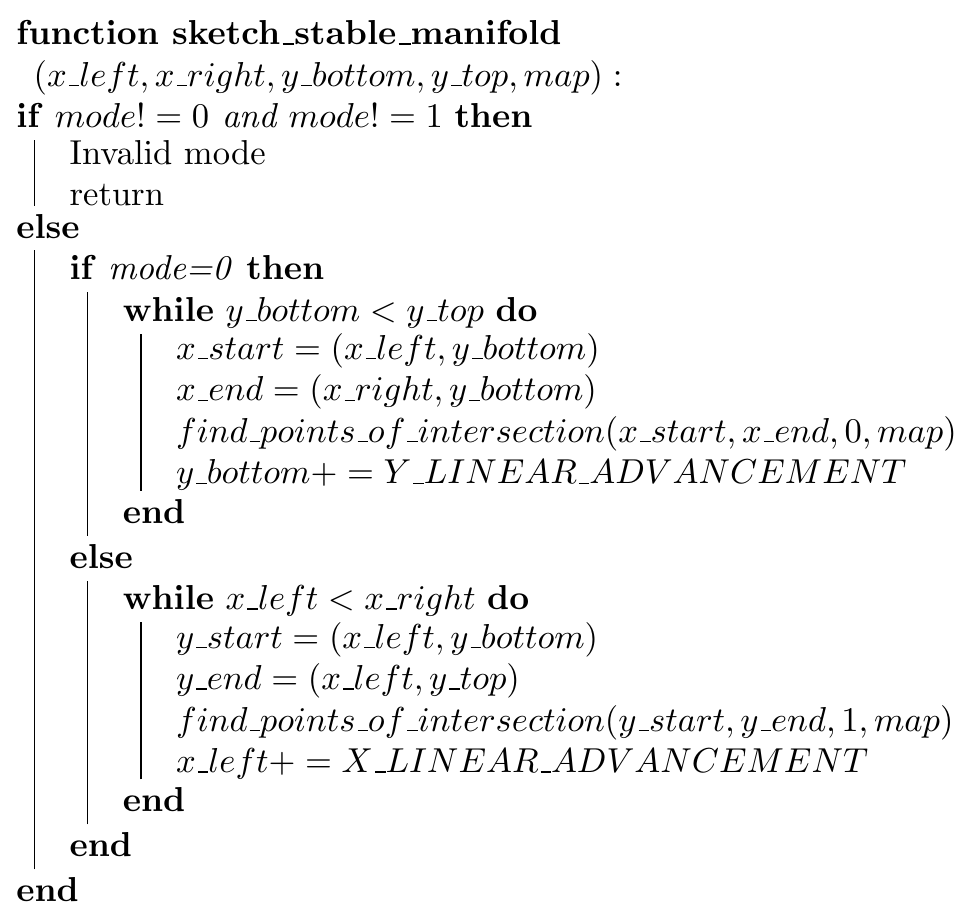} \\
        \hline        
    \end{tabular}
\end{table*}

\section{Conclusion}

In this paper we have introduced a new algorithm to plot the stable
manifolds of saddle points in 2-D maps. It does not require
specification of the inverse, nor does it depend on computation of
pre-iterates. The Point-Iterative algorithm identifies the points of
intersection of the stable manifold with the boundary of the region of
interest by utilizing the repelling property of the stable manifold. Forward iteration from these points give points on the stable manifold. We have described the implementation of the algorithm, and have shown that the algorithm can successfully plot the stable manifold in non-invertible maps even when the manifold has sharp folds and bends.

\section*{Acknowledgments} \noindent S.B. acknowledges the financial support from J C Bose Fellowship of the Science \& Engineering Research Board, Govt. of India, no. JBR/2020/000049.

\section*{Appendix} \noindent \label{Appendix}

\noindent Here, we provide the pseudo-code implementations of the functions necessary to sketch the stable manifold using the Point-Iterative Algorithm.


\end{multicols*}
\clearpage
\end{document}